# Enhancement of $T_C$ and reentrant spin-glass transition in $La_{0.86}Ca_{0.14}Mn_{1-y}Cr_yO_3$ (y = 0, 0.1 and 0.2)

T. Sudyoadsuk[1,2], R. Suryanarayanan[2*] and P. Winotai[1]

1. Department of Chemistry, Mahidol University, Rama VI Road, Bangkok 10400, Thailand
2. Laboratoire de Physico-Chimie de l'Etat Solide, UMR8648,Bât 414 Université Paris-Sud, 91405 Orsay, France

*corresponding author: ramanathan.suryan@lpces.u-psud.fr

## Abstract

We report on the structural, frequency dependent ac susceptibility, dc magnetization and magnetoresistance (MR) measurements on polycrystalline samples of $La_{0.86}Ca_{0.14}Mn_{1-y}Cr_yO_3$(y = 0, 0.1 and 0.2) prepared by sol-gel technique. For y = 0, a paramagnetic to ferromagnetic transition was observed at $T_C$ = 185 K. For y = 0.1, the value of $T_C$ = 200 K, an increase of 15 K and for y = 0.2, the $T_C$ = 195 K, an increase of 10 K. The imaginary part of the ac susceptibility of all the three samples shows a secondary transition at $T_f < T_C$. For y = 0, there is no definite law to account for the frequency dependence of $T_f$ and is attributed to a transition arising out of a canted structure. However, in the case of y = 0.1 and 0.2, the frequency dependence indicate the presence of a reentrant spin glass transition at $T_f$. Though all the three samples show a semi-conducting behavior between 300 and 5 K, a negative MR was observed corresponding to $T_C$ and $T_f$. The value of MR decreased for the Cr substituted samples.

Short title: *Reentrant spin glass transition in LCM(Cr)O*

## 1. Introduction

The manganites $Ln_{1-x}D_xMnO_3$ (Ln= Bi, rare earths; D= Pb, Ca, Sr, Ba) exhibit interesting physical properties such as paramagnetic to ferromagnetic transition (PFM), semi-conductor to metal transition (SMT), negative colossal magnetoresistance (CMR), charge ordering (CO), isotope effect etc and are being extensively investigated [1-4]. The divalent (D) substitution, and the dependence both on the value of x and on the ionic radius of the rare earth, introduces mixed valence of $Mn^{3+}$ and $Mn^{4+}$ in these compounds that results in the de-localization of $e_g$ electrons and a parallel alignment of $t_{2g}$ spins. This is the basis of the double exchange model which accounts for the occurrence of the paramagnetic to ferromagnetic (PM-FM) transition accompanied by a semi-conductor to metal transition (SMT) [5]. However, this model is inadequate to explain other properties as pointed out by several authors [3, 4, 6-8]. In particular, recent experiments seem to indicate that for certain values of x, the ground state is not a homogeneous ferromagnetic metallic (FMM) state but consists of regions of FM and anti-ferromagnetic (AFM) clusters [9]. The phase separation model relies heavily on such observations [3, 4]. Among manganites, the magnetic and magnetotransport properties of the $La_{1-x}Ca_xMnO_3$ compounds have been fairly well studied [1, 2]. A variety of magnetic transitions are observed as x increases from 0 to 1. Magnetization and neutron scattering data have been reported on single crystals of $La_{1-x}Ca_xMnO_3$ for x = 0.1, 0.125 and 0.2 [10]. We note, in particular, the AFM state that was present for x = 0, disappears for x = 0.1, resulting in a PM-FM transition at 138 K but the sample remains semi-conducting down to 5 K. In addition to the PM-FM transition, the sample shows a secondary transition at 112 K resulting from a canted spin structure as inferred from neutron measurements. The sample with x = 0.125 shows similar features. The SMT sets in only for x = 0.2 and the canted structure disappears. We note that no experimental data exists for x = 0.14, though one expects for this composition, from the phase diagram of Biotteau et al [10], a PM-FM transition accompanied by a canted structure at T not from 100 K. We are interested here in studying the effect of Cr substitution on the magnetic and magnetotransport properties of this lightly hole doped compound. Among the 3d-elements, Cr substitution is particularly interesting as $Cr^{3+}$ is iso-electronic with $Mn^{4+}$ and is a non-Jahn-Teller ion. In addition, the nature of the magnetic interaction between $Cr^{3+}$-O-$Mn^{3+}$ is known to favor ferromagnetism through superexchange interaction. Several interesting properties have indeed been reported on the effect of Cr substitution on the Mn site of the perovskite manganites. Raveau et al [11] showed that 5% of Cr substituted on the Mn site, in $Pr_{0.5}Ca_{0.5}MnO_3$, not only completely suppressed the CO transition but also induced a PM-FM transition that was accompanied by a semi-conductor to metal transition and the sample showed a CMR in the presence of a moderate magnetic field. In the case of $La_{0.46}Sr_{0.54}Mn_{1-y}Cr_yMnO_3$, a reentrant spin glass behavior was observed for y = 0.02 [12]. In the case of electron doped compounds such as $La_{0.7}Ca_{0.3}$ $Mn_{1-y}Cr_yMnO_3$, the CO that was observed for y = 0 at 260 K was completely suppressed in the y = 0.2 sample and though no SMT was observed, a sizeable CMR was reported due to spin dependent hopping [13]. In another electron-doped manganite system, $Sm_{1-x}Ca_xMnO_3$, the substitution of Ni, Co and Cr on the Mn site resulted in the disappearance of the CO and a CMR effect and ferromagnetic semi-metallic ground state were observed [14-16]. With these in mind, we have carried out structural, frequency dependent ac susceptibility, dc magnetization and magnetoresistance measurements on the polycrystalline samples of $La_{1-x}Ca_xMn_{1-y}Cr_yO_3$ (y=0, 0.1 and 0.2) prepared by sol-gel technique

## 2. Experimental details

The polycrystalline samples reported here were prepared by the well known sol-gel technique. Stoichiometric amounts of $La(NO_3)_3 \cdot 6H_2O$ (99%), $CaCO_3$ (99.5%), $Cr(NO_3)_3 \cdot 9H_2O$ (97%), and $Mn(CH_3COO)_2$ $4H_2O$ (99%) were dissolved in a dilute $HNO_3$ solution with citric acid and ethylene glycol used as the chelating agents. The mixed solution was then heated until a dark colored resin material was formed. The resin was subsequently fired at 773 K and 1173 K in air to decompose the organic residual. The resultant powder was then ground, palletized, and sintered at 1573 K for 12 h.

Room-temperature Powder X-ray diffraction (XRD) measurements on these samples were carried out with a Bruker D8 Advanced diffractometer in the Bragg-Brentano geometry using $CuK_\alpha$ radiation. The XRD patterns were recorded from 2θ = 20° to 130° with a 0.02° step size and a counting time of 12 sec. Further analysis by the Rietveld method using the *FULLPROF* program was carried out [17]. The surface morphology and composition of the samples was examined by a scanning electron microscope (SEM) fitted with an energy dispersive spectrometer. The SEM micrographs showed a granular structure for all the three samples with a highly dense morphology with little or no porosity and typical grain sizes on the order of 1 μm. The Energy Dispersive X-ray analysis indicated that the composition of the cations was quite close to the nominal one. Both the real (χ') and imaginary (χ'') parts of the ac susceptibility of the samples were measured as a function of temperature using a home built apparatus in which one could vary the frequency of the ac exciting field from 40 to 18000 Hz. The dc magnetization as a function of temperature and magnetic field was measured using a commercial SQUID magnetometer. Resistance was measured as function of temperature and in zero and in an applied field of 5 Tesla using a special sample holder that could be inserted into the cryostat of the SQUID magnetometer. Silver paint was used to make electrical contacts and the standard four point technique was used for these measurements.

## 3. Results

### 3.1 XRD data

The XRD analysis revealed neither the presence of any impurity phases nor any precipitates of unreacted oxides. These patterns could be indexed to the $GdFeO_3$-perovskite structure with the space group *Pbnm*. The calculated lattice parameters and other details are tabulated in Table I. The unit cell volume and the average distance $\langle d_{Mn\text{-}O} \rangle$ of the sample decreases as a function of Cr content. There is a general tendency for the angle $\theta_{<\text{Mn-O-Mn}>}$ to increase as the Cr intent increases though this increase is more notable for the y = 0.1 sample. The distortion of the $MnO_6$ octahedra in the 0.2 Cr-doped sample ($\sigma^2_{JT} = 2.92\times10^{-4}$) is much higher than that of the parent sample ($\sigma^2_{JT}$ $=1.27\times10^{-5}$)

3.2 ac susceptibility

The real and the imaginary parts of the *ac* susceptibility of the samples as a function of temperature measured in an ac excitation field of 5.5 Oe with a frequency set at 1.5 kHz, are shown in fig.1 (a, b, c). Several interesting features can be noticed. For y(Cr) = 0, χ' rises abruptly as T approaches 190 K, indicative of a paramagnetic to ferromagnetic transition (PM-FM), shows a small peak at 173 K and decreases slowly with a broad maximum around 120 K. On the other hand, χ" shows a step-like feature at ($T_C$) 185 K and rapidly increases with a flat region between 150 and 125 K followed by a somewhat broad maximum centered at ($T_f$) 95 K. For y(Cr)=0.1, there is a remarkable increase in the χ' peak from 173 to 189 K and the intensity decreases around 115 K. The step-like feature in χ" is also seen to occur at a higher temperature (200 K) indicating a net increase of about 15 K in $T_C$. The maximum in χ" also shifts to a higher temperature at $T_f$ = 108 K and the flat region seen in the case of y = 0 is somewhat suppressed indicating a change in magnetic structure in the region below 150 K. For a further increase in the Cr concentration to y = 0.2, both the peak in χ' and the step like feature in χ" shift to lower temperatures respectively to 180 and 195 K and the second maximum in χ" broadens centered at 100 K. With respect to y = 0, there is a net increase in $T_C$ of 10 K and further the flat region below 150 K resembles that observed in the case of y = 0.1.

We next demonstrate the effect of ac frequency of the exciting field on the imaginary part of the ac susceptibility of these samples. Though the low temperature part of χ' was affected by the change in frequency (*f*), it is the low temperature part of the χ" (showing a fairly well defined maximum) that exhibits easily measurable shifts. As a typical example, fig.2 (a, b, c) show the data at three different frequencies. The value of $T_f$ shifts to higher temperature as the frequency of the measurement increased. This phenomenon is characteristic of spin glasses [18]. To understand the nature of this transition, we have plotted $T_f$ as a function of *f* (fig.3). A straight line fit is obtained only in the case of Cr substituted samples indicating a different mechanism involved in the case of the parent compound(y=0). In the case of y = 0.1 and 0.2, we have determined the freezing temperature ($T_g$) by extrapolating the curve to zero frequency Taking the experimental values of $T_f$ and the extrapolated value of $T_g$ for these two samples, we have used the conventional spin-glass formula for critical slowing down to analyze our data (fig.4). According to this model [18]

$$\tau / \tau_0 = [\,(T_f - T_g) / T_g\,]^{-z\nu} \qquad (1)$$

Here, $T_g$ is the spin-glass transition temperature and $T_f$ is the frequency-dependent freezing temperature at which the maximum relaxation time τ corresponds to the measured frequency, $\tau_0$ is the characteristic time constant of the spin glass and zν is a critical component. The data for y=0.1 and 0.2 satisfy this relation yielding respectively $\tau_0 = 2.89 \times 10^{-9}$ s and $1.78 \times 10^{-8}$ s and zν = 6.05 and 6.17. (fig.4).

3.3 dc Magnetization

The magnetization (M) of the three samples at 6 K, as a function of magnetic field is shown in fig.5. The magnetization of all the samples are saturated at relatively low fields and the values of the saturated magnetic moment for y= 0, 0.1 and 0.2 are respectively 3.72, 3.09 and 2.41$\mu_B$/f.u. The zero field cooled (zfc) and field cooled (fc) magnetization measured in a field of 0.005 T, as a function of temperature, is shown in fig.6 (a, b, c). The data show clearly that the substitution of Cr enhances the value of $T_C$. Further, for all the samples, there is a difference between zfc and fc data. This may be indicative of spin glass like behavior or merely indicate irreversible effects which can be due to irreversible domain wall dynamics. The difference between zfc and fc at low temperatures decreases as a function of Cr substitution indicating a modification of magnetic structure. This difference almost vanishes for H = 1 T for all the three samples. We have also tried to fit the high temperature (200 < *T* < 300 K) magnetization data to Curie-Weiss law (not shown here). The data yielded a positive intercept at 194, 215 and 198 K respectively for y = 0, 0.1 and 0.2 and the respective moments (H=0.005 T) $p_{eff}$ were 6.18, 5.32 and 4.5 $\mu_B$. The values of $p_{eff}$ were found to depend on the applied magnetic field and in general increased with the field. Thus, these values for the three samples were 6.53 and 6.63 $\mu_B$; 5.54and 5.67 $\mu_B$; 4.78 and 5.02 $\mu_B$ when the applied field was 0.05 and 1 T respectively.

3.4 Resistivity and magnetoresistance

Resistivity ($\rho$) of the three samples, as a function of temperature in zero ($\rho_0$) and 5T ($\rho_H$) are shown in fig.7. All the three samples show semi-conducting behavior in zero field. The data show a small change in slope at temperatures corresponding to $T_c$; Though the samples remain semi-conductors in an applied field of 5 T, the change in slope is suppressed and the resistivity decreases especially for the samples y = 0 and y = 0.1. The magneto resistance [MR % = -100 x ($\rho_H$ - $\rho_0$)/ $\rho_0$] shown as a function of temperature in fig.8 reveals interesting features. For all the three samples, a non zero MR is observed close to 300 K which is much higher than the $T_c$. For y = 0, the MR increases rapidly from near 300 K reaching values greater than 50% and shows two distinct maxima at 180 and at 100 K. The high resistivity for $T$ < 50 K did not allow us to obtain reliable data on MR, though MR showed an increase for $T$ > 75 K. For y = 0.1, the value of MR reduced to 30% but a distinct maximum in MR was observed at 200 K followed by a broad maximum centered at 125 K. For y = 0.2, the value of MR was further reduced to about 20% and only a broad maximum centered at around 160 K was observed.

4. Discussion

It is well known that in the manganites, $Ln_{1-x}D_xMnO_3$, there is a strong competition between superexchange and double exchange interactions. This leads to a variety of magnetic orderings depending on the ionic radius of Ln and D and the value of x. The samples of our interest are in the lightly hole doped region of the phase diagram of $La_{1-x}Ca_xMnO_3$. Though, Biotteau et al. [10] have not reported any data for x=0.14, the composition of our interest, it is reasonable to extrapolate the properties from the published phase diagram. Thus, one would expect that the sample with x = 0.14 shows a PM-FM transition followed by another transition with a canted structure similar to that reported for x = 0.1 and 0.125. Indeed, our ac susceptibility data indicate a PM-FM transition at $T_c \approx$ 185 K with another secondary transition at a lower temperature $T_f$ =95 K. The frequency response of this secondary transition does not provide a good fit with eqn(1) to identify it with a conventional spin glass. One may attribute this to a canted structure similar to that observed for x=0.125 by Biotteau et al[10] though neutron scattering experiments have to be carried out to ascertain this. Following, Dagotto et al. [4], this transition may arise from spin cluster glass, an anti-ferromagnetic matrix containing ferromagnetic clusters, that can give rise to a magntoresistance effect as was indeed observed in our sample. As pointed out by Dagotto [19] the exact nature of these spin clusters vis a vis the classical spin glass is not clear at the present moment and is currently under investigation by several others. We wish to add further that the sample x=0, reported to exhibit a canted spin structure by Biotteau et al[10], was reexamined by Yates et al. [20] by using NMR techniques. Their data do not rule out the possibility that the ferromagnetic host is canted but certainly do not support the case for true long range phase coexistence and thus indicating the complexity of the properties of these materials.

The substitution of Cr leads to three noticeable effects- an increase in $T_C$, a reentrant spin glass transition at lower temperatures and a suppression of CMR. The increase in $T_C$ can be related to the increase of the FM exchange interaction between $Mn^{3+}$ and $Cr^{3+}$ as can be inferred from the increase in the paramagnetic Curie-Weiss temperature and also to the change in structural parameters like $\theta_{<Mn\text{-}O\text{-}Mn>}$. The presence of $Cr^{3+}$ effectively contributes to an increase in $Mn^{4+}$ content and can account for the reduction in the saturated magnetic moment. On the other hand, the high value of the $p_{eff}$ compared to the theoretically expected value and the dependence on the magnetic field is attributed to the presence of magnetic polarons. Such a behavior was also observed by Martinez et al., [21] in the case of $La_{0.9}Ca_{0.1}MnO_3$ . The $p_{eff}$ of a single crystal of this composition increased from 4.8 to 5.8 $\mu_B$/f.u. when the measured field was increased from 2.5 x $10^{-4}$ T to 5 T.

The transitions ($\chi$'')observed at $T_f$ for the y = 0.1 and 0.2 samples exhibit a frequency dependence that is well known in the case of a spin glass. In the case of a conventional spin glass one observes $\tau_0 = 10^{-13}$ s whereas the value obtained here is somewhat larger, $10^{-10}$ s. Such a difference may be attributed to the presence of ferromagnetic clusters. Several others have observed either reentrant spin glass behavior or spin glass transition in manganites and cobaltates. We will present a short discussion of a few cases that are relevant to the present study. Studies on manganites exhibiting spin-glass behavior display varying degrees of magnetoresistance ranging from a CMR in $(La_{1/3}\ Tb_{2/3})_{2/3}Ca_{1/3}MnO_3$ as a result of a field-induced semiconductor-to-metal transition for fields greater than 5 T [22] to no magnetoresistance effect in $Y_{0.5}Sr_{0.5}MnO_3$ for fields as high as 4 T [23]. The compound $La_{0.46}Sr_{0.56}Mn_{1-y}Cr_yO_3$ shows a PM to FM and FM to AFM transitions at 272 and 190 K respectively [12]. For y = 0.02, in addition to the AFM transition, these authors observed a reentrant spin glass transition at lower temperatures and their data follow the Eqn.(1) to yield a value of $\tau_0 = 10^{-13}$ s. However, no CMR data were reported. In the case of the compound $La_{0.95}Sr_{0.05}CoO_3$, only one magnetic transition was identified at 15 K which showed a frequency dependent ac susceptibility yielding a value of $\tau_0 = 2.3 \times 10^{-10}$ s [24]. Maignan et al. [25] examined the magnetic properties of the compounds $Th_{0.35}D_{0.65}MnO_3$ and showed the role played by the ionic size of D on the appearance of spin glass insulating state.

Another interesting system is $La_{0.96-z}Nd_zK_{0.04}MnO_3$ in which a disordered magnetic state is formed as La is substituted by Nd(z), reflecting the competition between FM double exchange and AFM superexchange interactions [26]. In particular, for z = 0.4, a reentrant spin glass transition was reported at $T_f$ ($T_f < T_C$) and further in a field of 5 T a second maximum in the MR was noticed corresponding to $T_f$ quite similar to what we report here. In our case, already for y = 0, the system is in a disordered state as indicated by an additional transition (attributed to a canted structure) at $T_f$ ($T_f < T_C$). A moderate field of 5 T is able to reduce the spin disorder scattering resulting in a MR effect centered at $T_f$. The substitution of Cr further causes a change in the magnetic structure near $T_f$ resulting in an increased frustration. This would increase the zero field resistivity at low temperatures and reduce the MR effect as observed. On the other hand, the MR>0 for $T>T_c$ which may be attributed to the presence of magnetic polarons as was also suggested by Martinez et al. [21] in the case of $La_{0.9}Ca_{0.1}MnO_3$ .

We also need to consider other points such as (a) the impact of charge ordering and its stability under Cr doping and (b) orbital ordering. For example, for compositions x>0.14, Dai et al [27] have pointed out that charge ordering becomes short range. And, Kimura et al [28] have shown that charge ordering can be destroyed to give a ferromagnetic ordering by Cr doping. Hence, to account for the increase in $T_c$, one might propose that an additional contribution may arise form such a destruction of charge ordering even on local scale. Regarding the point (b), one may note that with various dopings, orbital ordering can be frustrated to produce a Jahn-Teller glass, that is a frustrated glassy like arrangement of $e_g$ orbital. In fact, in $Pr_{0.7}Ca_{0.3}MnO_3$, such states were shown to behave like a spin glass [29].

We would like to conclude this section by commenting briefly on two models that are relevant to the present work. The first one is the phase separation model [3,4] according to which the competition between AFM and FM states leads to inhomogeneities, especially in the case of lightly doped manganites, resulting in phase-segregation. This phase-segregated state may have some similarities to the classical spin glass similar to what we have reported here. An increased presence of the AFM component does not lead to the metallic state in zero magnetic field but a moderate field would enhance the tunneling due to field induced alignment of magnetic domains resulting in a negative MR. The second, recent, model considers three strong on-site interactions, namely Jahn-Teller (JT) coupling which splits the two fold $e_g$ orbital degeneracy, Hund's coupling and $e_g$ electron repulsion [8]. The new approach by these authors is to consider the coexistence of JT polaronic (*l* band) and the broad band of $e_g$ states (*b* band). According to these authors, the effective band width 2D of the *b* electrons decrease (for small Ca concentration) and the *b* band bottom is above the *l* level. This leads to a low temperature insulating state. Further, the sample exhibits a PM-FM transition because of strong nearest neighbor exchange between $t_{2g}$ spins as is indeed observed in the present study. However, the details of a further canted or spin-glass like structure below $T_C$ have not been worked out yet in this model.

## 5. Conclusions

We have carried out structural, frequency dependent ac susceptibility, dc magnetization and magnetoresistance (MR) measurements on polycrystalline samples of $La_{0.86}Ca_{0.14}Mn_{1-y}Cr_yO_3$ (y = 0, 0.1 and 0.2) prepared by sol-gel technique. For y = 0, a paramagnetic to ferromagnetic transition was observed at $T_C$ = 185 K. The Cr substitution increases the value of $T_C$ to 200 K for y = 0.1 and to 195 K for y = 0.2. The imaginary part of the ac susceptibility shows a secondary transition at $T_f < T_C$ for y = 0, which is attributed to a transition arising out of a canted structure. However, in the case of y = 0.1 and 0.2, the frequency dependence indicate the presence of a reentrant spin glass transition at $T_f$. Though all the three samples do not show a semi-conductor to metal transition between 300 and 5 K, a negative MR was observed corresponding to $T_C$ and $T_f$. The value of MR decreased for the Cr substituted samples. Since the spin glass phase observed here giving rise to a CMR effect might have some similarities with the phase segregated states evoked in certain models [3,4,19], we believe that the present work would be of considerable significance.

Acknowledgments

TS, PW, and RS thank the Thai Royal Golden Jubilee program for financial support and to PERCH for partial financial support. TS wishes to thank A. Revcolevschi for the hospitality shown to him during his stay in Orsay.

References

[1] Rao C N R and Raveau B (ed) 1998 *Colossal Magnetoresistance, Charge Ordering and Related Properties of Manganese Oxides* (World Scientific, Singapore).
[2] Tokura Y (ed) 2000 *Colossal Magnetoresistive Oxides* (Gordon and Breach, New York).
[3] Nagaev E L 2002 *Colossal Magnetoresistance and Phase Separation in Magnetic Semiconductors* (Imperial College Press, London).
[4] Dagotto E, Hotta T, and Moreo A 2001 Phys. Reports **344** 1.
[5] Zener C 1951 Phys. Rev. **82** 403.
[6] Millis A J, Shraiman B I, and Littlewood P B 1995 Phys. Rev. Lett. **74** 5144.
[7] Millis A J, Shraiman B I, and Mueller R 1996 Phys. Rev. B **54** 5389.
[8] Venketeswara Pai G, Hassan S R, Krishnamurthy H R and Ramakrishnan T V 2003, Europhysics Lett. **64** 696; Ramakrishnan T V, Krishnamurthy H R, Hassan S R and Venketeswara Pai G in Chatterji T (ed) 2003 *Colossal Magnetoresistive Manganites* (Kluwer Academic Publishers, Dordecht, The Netherlands) and also arXiv:cond-mat/038376.
[9] Fäth M, Freisem S, Menovsky A A, Tomioka Y Arts J and Mydosh J A 1999 Science **285** 1540; Uehara M, Mori S, Chen C H and Cheong S W 1999 Nature **399** 560
[10] Biotteau G, Hennion M, Mossa F, Rodriguez-Caravajal, Pinsard L and Revcolevschi A 2001 Phys. Rev. B **64** 104421.
[11] Raveau B, Maignan A, and Martin C 1997 J. Solid State Chem.**130** 162
[12] Dho J, Kim W S and Hor N H 2002 Phys. Rev. Lett. **89** 027202
[13] Sudadyosuk T, Suryanarayanan R and Winotai P JMMM(in press)
[14] Damay F, Martin C, Maignan A, and Raveau B 1998 J. Mag. Mag. Mater. **183** 143.
[15] Maignan A, Martin C, Damay F, Hervieu M, and Raveau B 1998 J. Mag. Mag. Mater. **188** 185.
[16] Maignan A ,Martin C, and Raveau B 1999 Mater. Res. Bull. **34** 345.
[17] Rodriguez-Carvajal J, 1993 Physica B **192** 55.
[18] Mydosh J A 1993 *Spin glasses- An introduction* (Taylor and Francis, London,UK)
[19] Dagotto E 2002 *Nanoscale Phase Separation and Colossal Magnetoresistance* (Springer-Verlag, Berlin), Chapter16.
[20] Yates K A, Kapusta C, Riedi P C, Ghivelder L and Cohen L F 2003 J. Mag. Mag. Mater. **260** 105
[21] Martinez B, Laukhin V, Fontcuberta J, Pinsard L and Revcolevschi A 2002 Phys. Rev. B **66** 054436.
[22] de Teresa M D, Ibarra M R, García J, Blasco J, Ritter C, Algarabel P A, Marquina C, and del Moral A 1996 Phys. Rev. Lett. **76** 3392.
[23] Chatterjee S and Nigam A K, 2002 Phys. Rev. B **66** 104403.
[24] Nam D N H, Mathieu R, Nordblad R, Khiem N V and Phuc N X 2000 Phys. Rev. B **62** 8989.
[25] Maignan A, Martin C, Van Tendeloo G, Hervieu M and Raveau B 1999 Phys. Rev. B **60** 15214.
[26] Mathieu R, Svendlindh P and Nordblad P 2000 Europhysics Lett.. **52** 441.
[27] Dai P, Fernandez-Baca J A, Wakabayashi N, Plumer E W, Tomioka Y and Tokura Y 2000 Phys. Rev. Lett. **85** 2553
[28] Kimura T, Kumai R, Okimoto Y, Tomioka Y and Tokura Y 2000 Phys. Rev. B **62** 15021
[29] Radaelli P G, Ibarra R M, Argyriou D N, Casalta H, Andersen K M, Cheong S W and Mitchell J F 2001 Phys. Rev. B **63** 172419

Figure Captions

Figure 1. Real ($\chi$') and imaginary ($\chi$'') parts of the a susceptibility of $La_{0.86}Ca_{0.14}Mn_{1-y}Cr_yO_3$ as a function of temperature.(a) y=0; (b)y=0.1; (c) y=0.2.

Figure 2. Imaginary part ($\chi$'') of the ac susceptibility of $La_{0.86}Ca_{0.14}Mn_{1-y}Cr_yO_3$ as a function of temperature at three different frequencies. (a) y=0; (b)y=0.1; (c) y=0.2.

Figure 3. $T_f$ of $La_{0.86}Ca_{0.14}Mn_{1-y}Cr_yO_3$ (y=0.1 and 0.2) as a function of frequency (*f*).

Figure 4. ln $\tau$ as a function of ln [ $(T_f - T_g)/ T_g$ ].(a) y=0; (b) y=0.2.

Figure 5. Magnetization at 5 K of $La_{0.86}Ca_{0.14}Mn_{1-y}Cr_yO_3$ (y=0, 0.1 and 0.2) as a function of field

Figure 6. Magnetization of $La_{0.86}Ca_{0.14}Mn_{1-y}Cr_yO_3$ as a function of temperature. (a) y=0; (b) y=0.1; (c) y= .2.

Figure 7. Resistivity (in H=0 and 5 T)of $La_{0.86}Ca_{0.14}Mn_{1-y}Cr_yO_3$ (y=0, 0.1 and 0.2) as a function of temperature. The data for the sample y=0.2 in H=5 T cannot be seen in the log scale because of a small magnetoreistance effect.

Figure 8. Magnetoresistance(see text for definition) in a field of 5 T of $La_{0.86}Ca_{0.14}Mn_{1-y}Cr_yO_3$ (y=0, 0.1 and 0.2) as a function of temperature.

**Table I.** Lattice parameters for $La_{0.86}Ca_{0.14}MnO_3$, $La_{0.86}Ca_{0.14}Mn_{0.90}Cr_{0.10}O_3$ and $La_{0.86}Ca_{0.14}Mn_{0.80}Cr_{0.20}O_3$ from the Rietveld refinement of X-ray diffraction data. The numbers in parentheses are estimated standard deviations to the last significant digit, and $\sigma^2_{JT} = 1/6\ \Sigma\{[d_{(\mathrm{Mn-O})i} - d_{<Mn-O>}\}/d_{<Mn-O>}]^2$.

| | $La_{0.86}Ca_{0.14}MnO_3$ | $La_{0.86}Ca_{0.14}Mn_{0.90}Cr_{0.10}O_3$ | $La_{0.86}Ca_{0.14}Mn_{0.80}Cr_{0.20}O_3$ |
|---|---|---|---|
| *a* (Å) | 5.4959(1) | 5.4987(1) | 5.4858(1) |
| *b* (Å) | 5.5040(1) | 5.5116(1) | 5.5057(1) |
| *c* (Å) | 7.7920(1) | 7.7713(2) | 7.7619(2) |
| *V* (Å$^3$) | 235.70(1) | 235.52(1) | 234.43(1) |
| $<d_{Mn-O}>$ | 1.975 | 1.967 | 1.964 |
| $\theta_{<Mn-O-Mn>}$ | 160.12° | 164.04° | 164.81° |
| $\sigma^2_{JT}$ | $1.27 \times 10^{-5}$ | $1.53 \times 10^{-5}$ | $2.92 \times 10^{-4}$ |

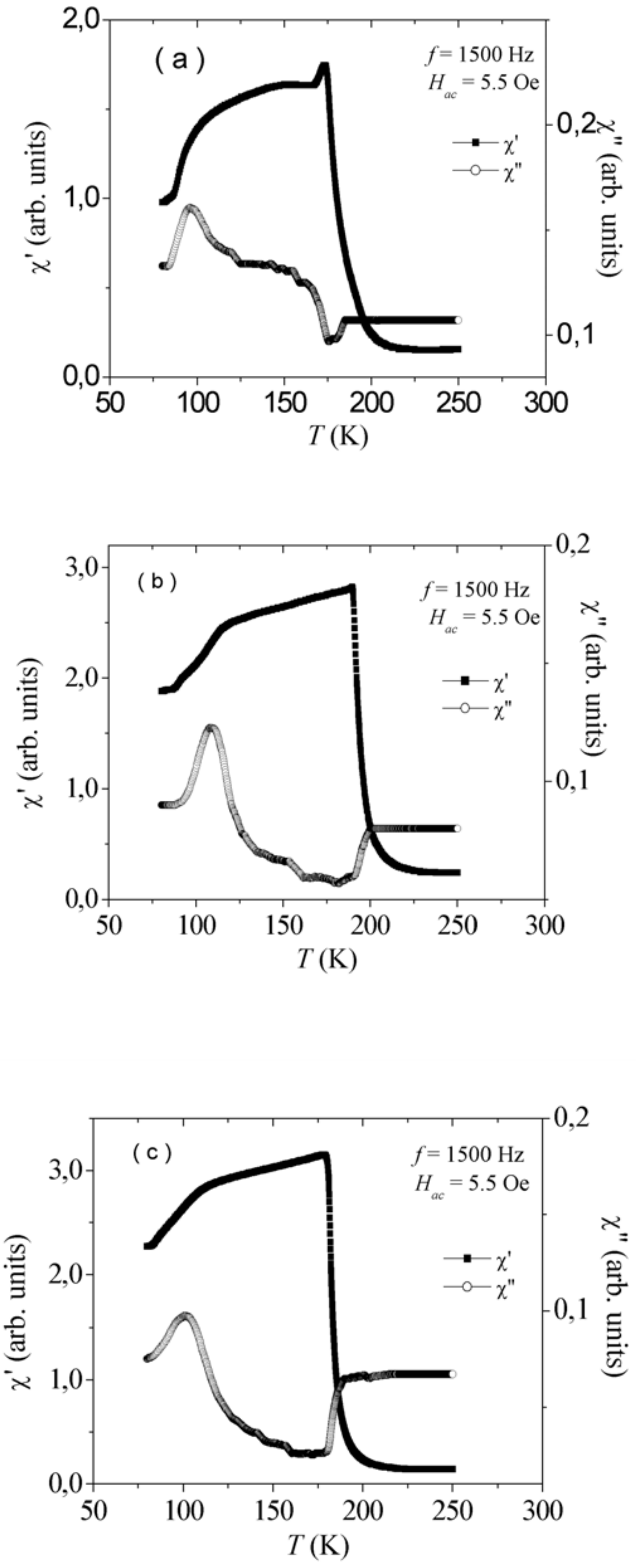
(a)
f = 1500 Hz
$H_{ac}$ = 5.5 Oe
χ'
χ"
χ' (arb. units)
χ" (arb. units)
T (K)
(b)
f = 1500 Hz
$H_{ac}$ = 5.5 Oe
χ'
χ"
χ' (arb. units)
χ" (arb. units)
T (K)
(c)
f = 1500 Hz
$H_{ac}$ = 5.5 Oe
χ'
χ"
χ' (arb. units)
χ" (arb. units)
T (K)

Fig 1 (a,b,c)

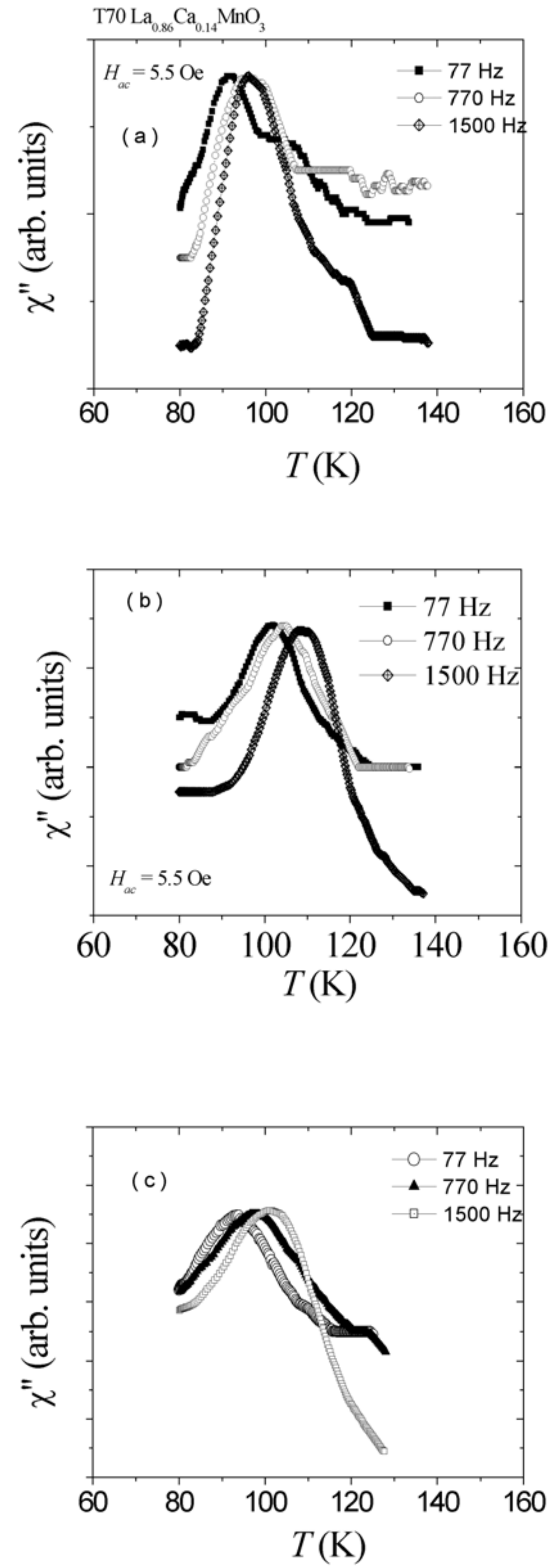
T70 $La_{0.86}Ca_{0.14}MnO_3$
$H_{ac}$ = 5.5 Oe
(a)
77 Hz
770 Hz
1500 Hz
χ'' (arb. units)
60 80 100 120 140 160
T (K)
(b)
77 Hz
770 Hz
1500 Hz
$H_{ac}$ = 5.5 Oe
χ'' (arb. units)
60 80 100 120 140 160
T (K)
(c)
77 Hz
770 Hz
1500 Hz
χ'' (arb. units)
60 80 100 120 140 160
T (K)

Fig. 2 (a,b,c)

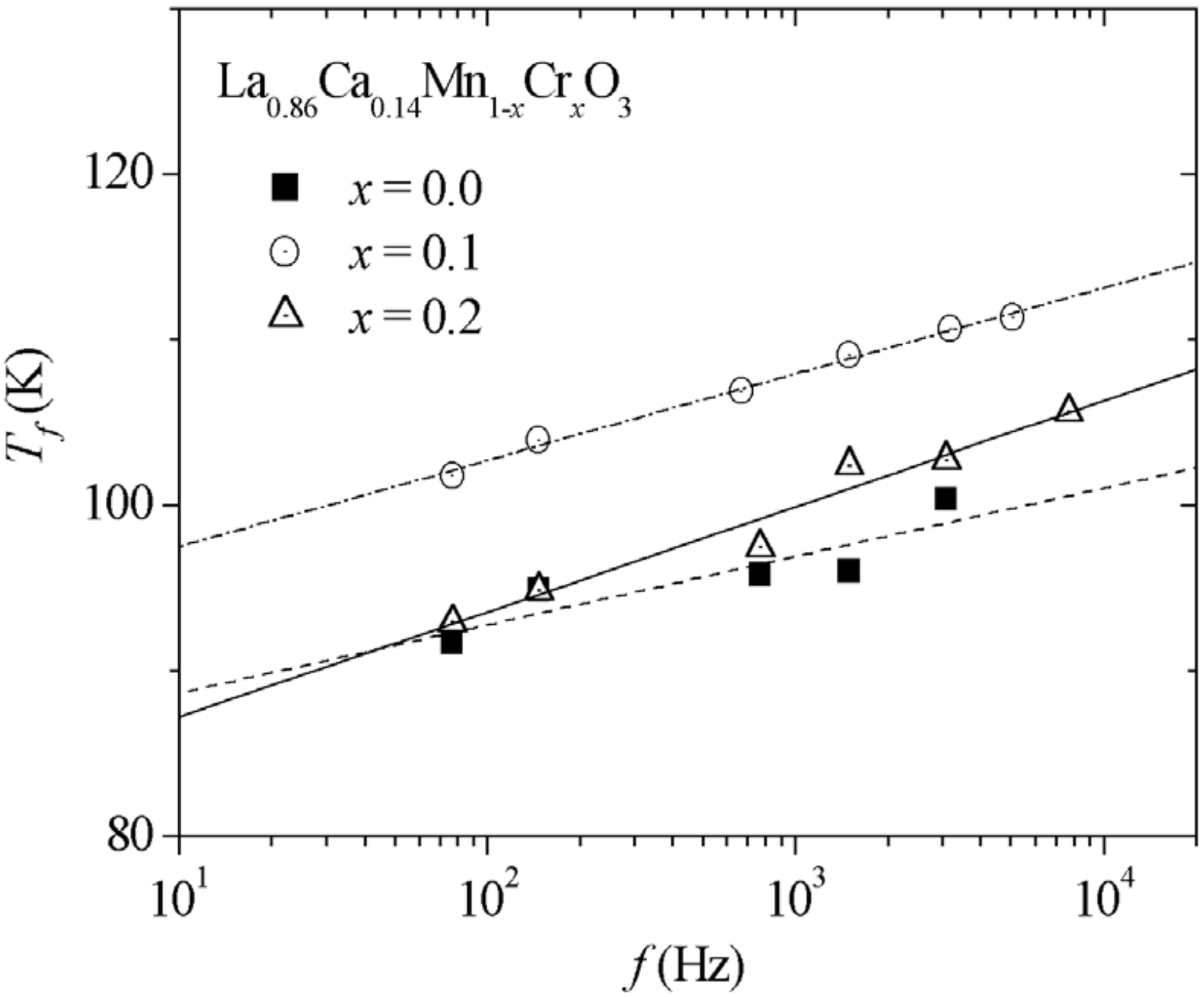

Fig. 3

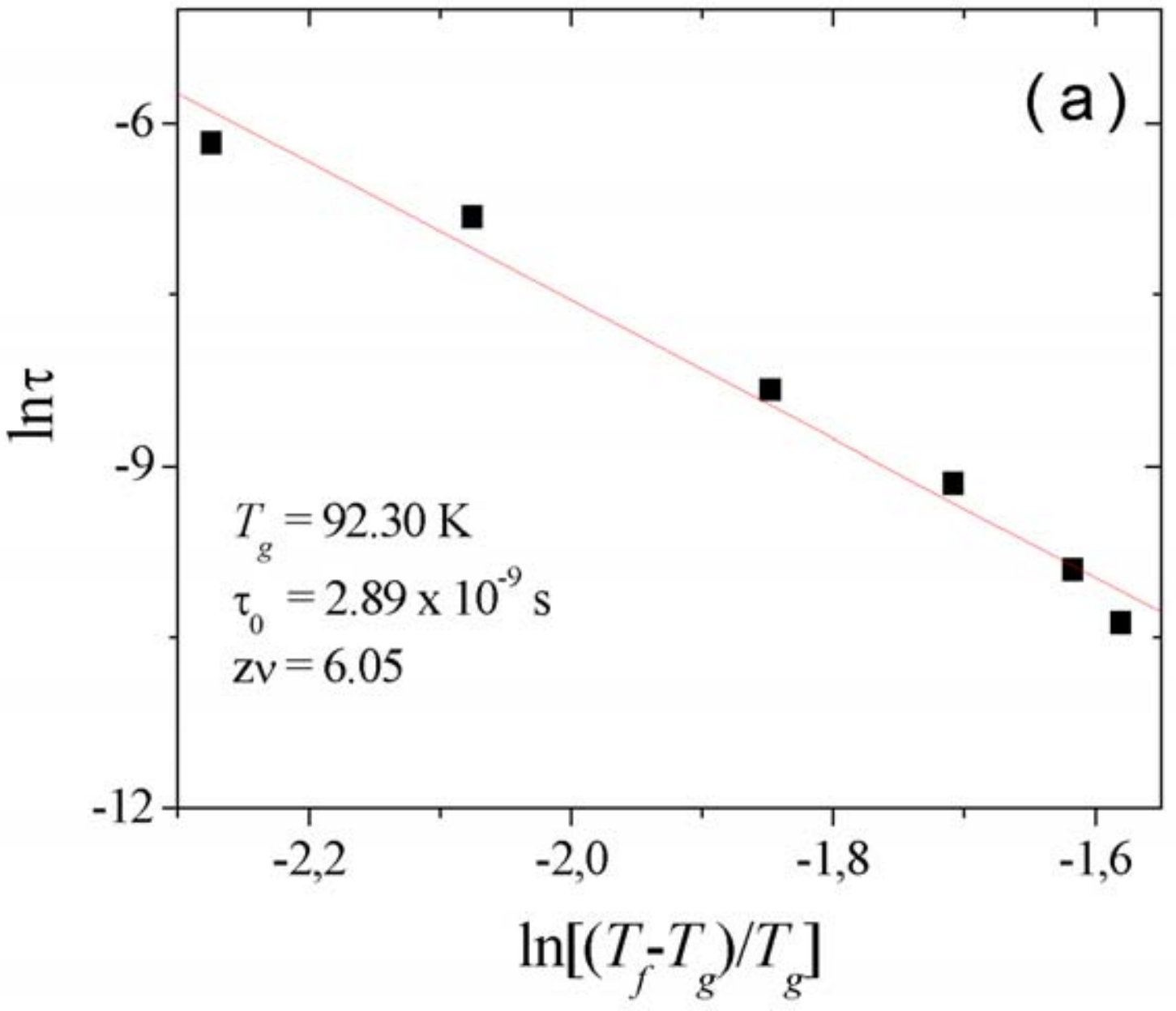

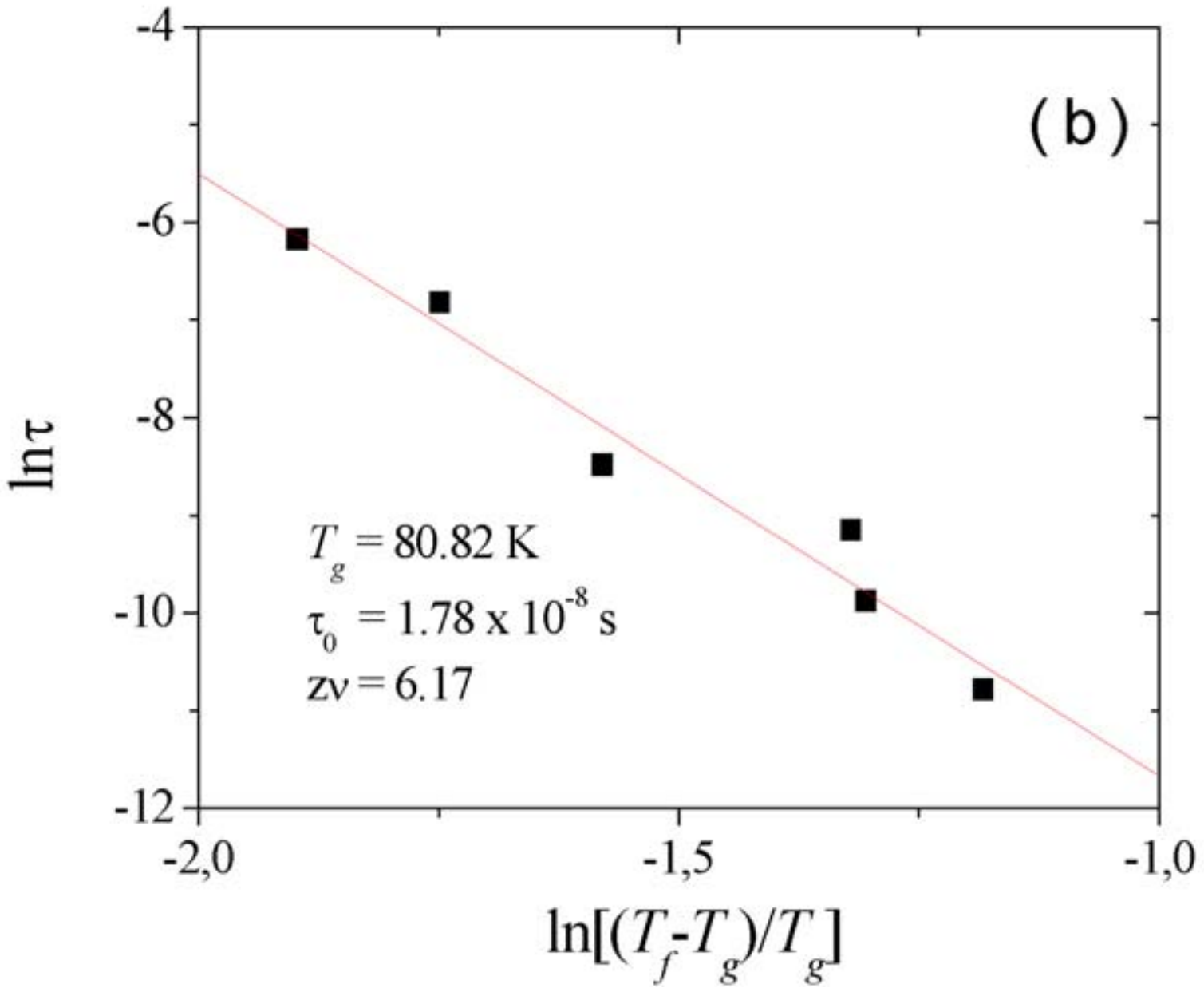

Fig.4(a,b)

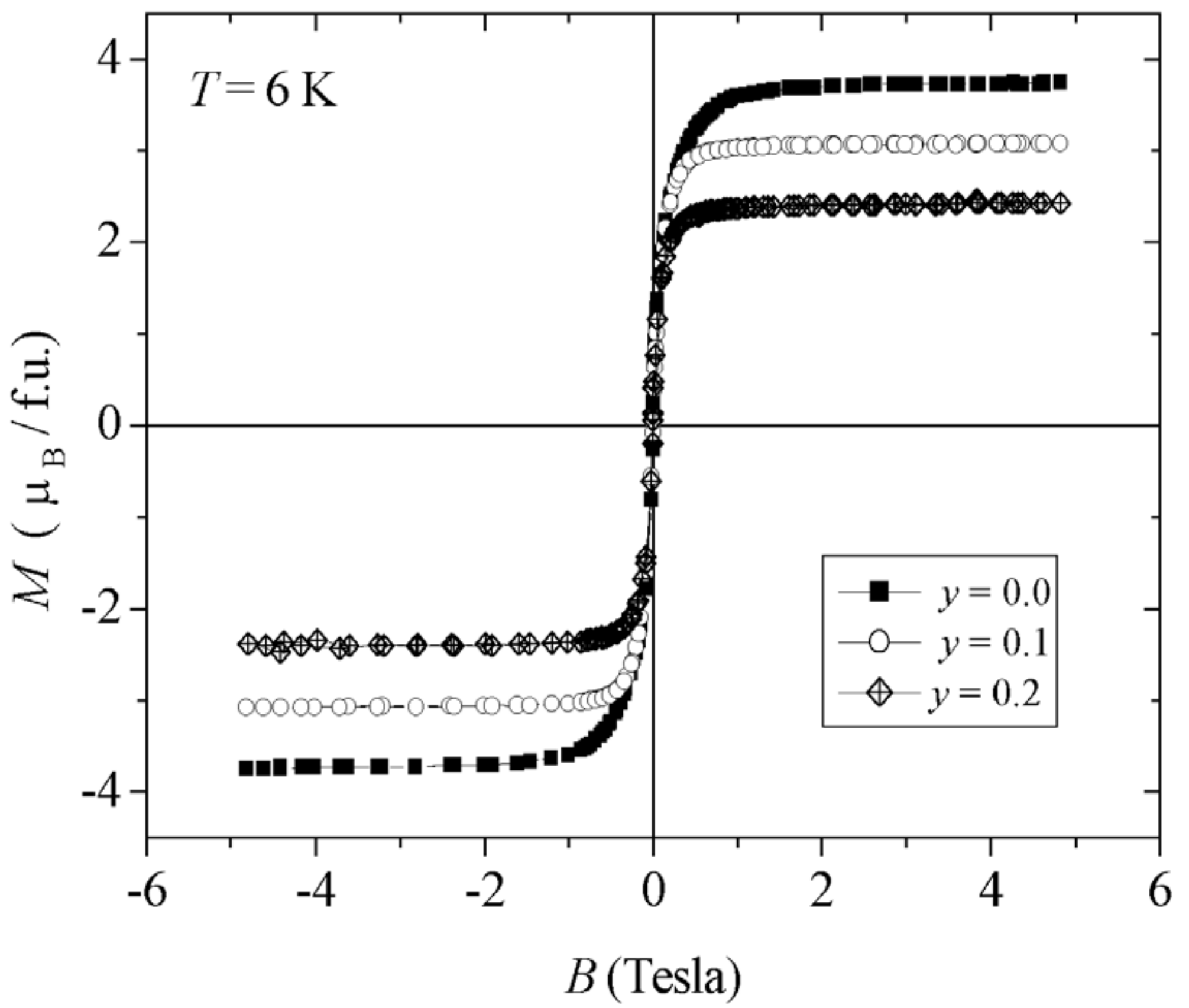

Fig. 5

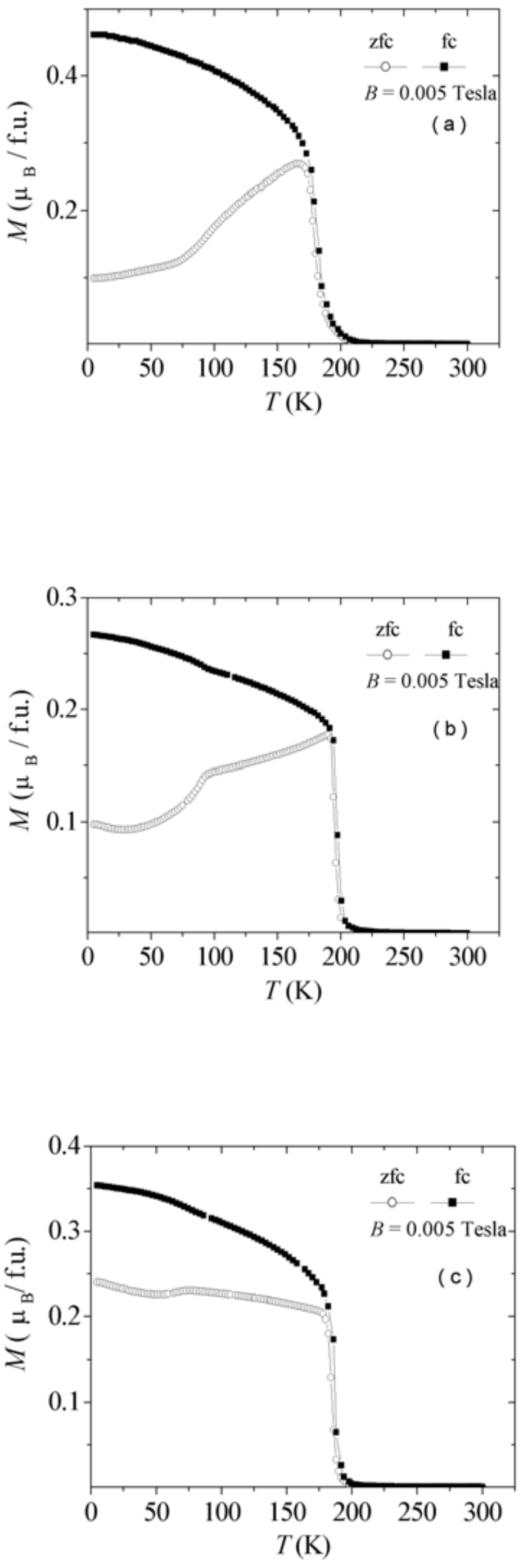
zfc
fc
B = 0.005 Tesla
( a )
M ( μ B / f.u.)
T (K)
zfc
fc
B = 0.005 Tesla
( b )
M ( μ B / f.u.)
T (K)
zfc
fc
B = 0.005 Tesla
( c )
M ( μB/ f.u.)
T (K)

Fig 6(a, b, c)

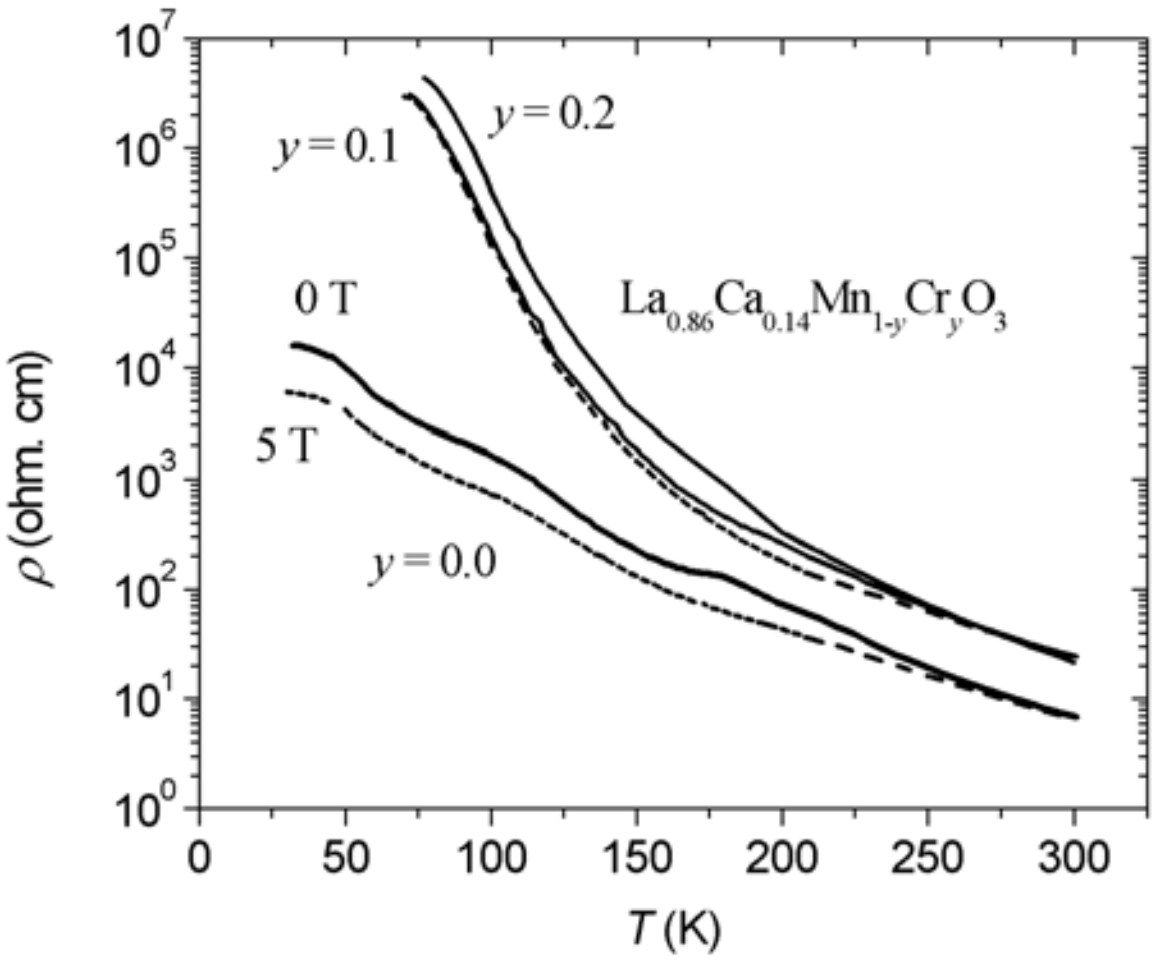

Fig. 7

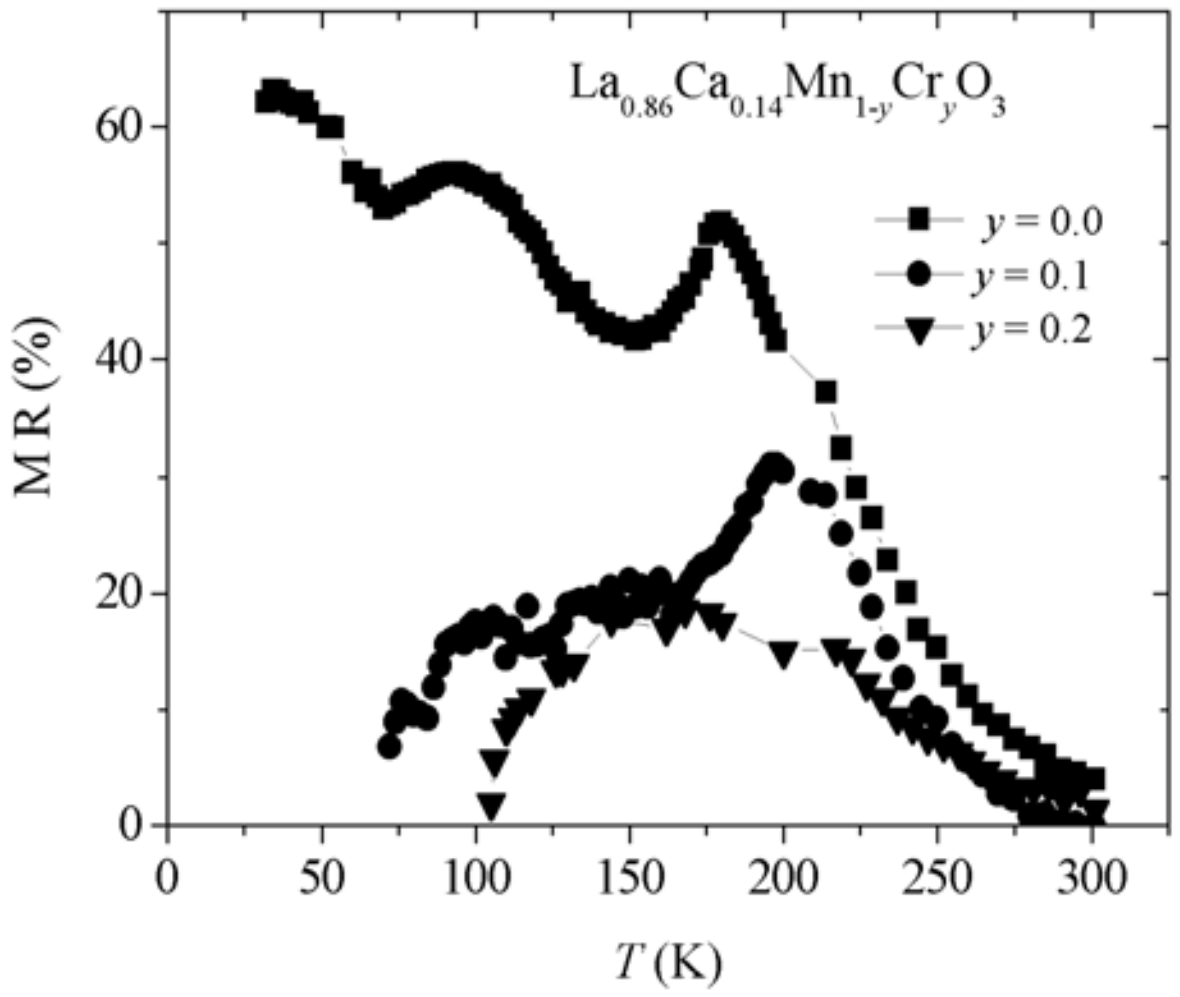

Fig.8